\documentclass[10pt]{article}
\usepackage{latexsym}
\usepackage{amssymb}
\usepackage{amsmath}
\usepackage{amscd}
\usepackage{amsthm}
\usepackage[left=2.5cm,top=2.75cm,right=2.5cm,bottom=1.75cm]{geometry}
\usepackage[dvips]{graphicx}
\usepackage{hyperref}
\begin{document}
\begin{center}
\large{\bf {Bianchi-V string cosmological model and late time acceleration}} \\
\vspace{10mm}
\normalsize{Anil Kumar Yadav}\\
\vspace{4mm}
\normalsize{Department of Physics, Anand Engineering
College, Keetham, Agra - 282 007, India} \\
\vspace{2mm}
\normalsize{E-mail: abanilyadav@yahoo.co.in}\\
\end{center}
\vspace{10mm}
\begin{abstract} 
In the present work we have searched the existence of the late time acceleration 
of the universe with string fluid as source of matter in Bianchi V space-time. To get the deterministic 
solution, we choose the scale factor as increasing function of time that yields a 
time dependent deceleration parameter, representing a model which generates 
a transition of universe from early decelerating phase to recent accelerating phase. 
The study reveals that strings dominate the early universe and eventually disappear from the 
universe for sufficiently large times i. e. at present epoch. The same is observed by current astronomical observations. 
The physical behaviour of universe has been discussed in detail.\\      
\end{abstract}
\section{Introduction}
In recent years, Einstein's theory of gravity has been subject of intense study for 
his success in explaining the observed accelerated expansion of late time universe. 
The substantial theoretical progress in string theory has brought
forth a diverse new generation of cosmological models, some of which are subject to
direct observational tests. 
The present day observations of universe indicate the existence of large scale 
network of strings in early universe (Kible, 1976, 1980). One key advance is the emergence of methods of moduli stabilization. 
Compactification of string theory from the total dimension D down to four dimensions introduces many 
gravitationally-coupled scalar fields moduli from the point of view of the four dimensional theory. 
Recently we have studied inhomogeneous string cosmological model
formed by geometric string and use this model as a source of gravitational field 
(Pradhan et al. 2007; Yadav et al. 2009). We had two main reason 
to study the above mentioned model. First, as a test of consistency, for some particular field theories based on string 
models and second we point out the universe can be represented by a collection of extended galaxies. It is generally 
assumed that after the big bang, the universe may have undergone a series of phase transitions as its temperature 
cooled below some critical 
temperature as predicted by grand unified theories (Zel'dovich et al. 1975; Kibble 1976, 1980; 
Everett 1981; Vilenkin 1981). At the very early stage of evolution 
of universe, it is believed that during the phase transition, the 
symmetry of universe was broken spontaneously. That could have given rise to topologically-stable defects such as 
domain walls, strings and monopoles (Vilenkin 1981). Among all the three cosmological structures, 
only cosmic strings have excited 
the most interesting consequence (Vilenkin 1985), because it gives rise the density perturbations which leads 
to the formation of galaxies. The cosmic string are important in the early stage of evolution 
of the universe before the particle creation because cosmic strings are one dimensional topological defects 
associated with spontaneous symmetry breaking whose plausible production site is cosmological phase transition in the 
early universe only. Also the present day observations reveal that the cosmic strings are not 
responsible for either the CMB fluctuations or the observed clustering of galaxies 
(Pogosian et al. 2006; Pogosian et al. 2003). The gravitational effect of cosmic strings have been extensively 
discussed by Letelier (1979) who considered the massive string to be 
formed by geometric strings with particle attached along its extension. Later on Letelier (1983) used this 
idea in obtaining cosmological solution in Bianchi-I and Kantowski-Sachs space-time.\\  

Recently, Pradhan and Amirhashchi (2011) have studied the law of variation of scale factor as increasing 
function of time in Bianchi-V space-time, which generates a time dependent deceleration parameter (DP). 
This law provides explicit 
form of scale factors governing the Bianchi-V universe and facilitates to describe 
the transition of universe from early decelerating phase to recent accelerating phase. Yadav et al. (2011) and 
Bali (2008) have obtained Bianchi-V string cosmological models in general relativity. 
The string cosmological models with the magnetic field are discussed by Chakraborty (1980), Tikekar and Patel 
(1992, 1994), Patel and Maharaj (1996). Singh and Singh (1999) investigated string cosmological model with 
magnetic field in the context of space-time with $G_{3}$ symmetry. Saha and Visineusu (2008) have studied 
string cosmological model in presence of magnetic flux and concluded that the presence of cosmic 
string does not allow the anisotropic universe to evolves into an isotropic one. Recently Yadav (2011) 
studied Bianchi-I string cosmological model with variable DP. The study of 
Bianchi-V cosmological model create more interest as these models contain isotropic special cases 
and permit arbitrary small anisotropy levels at some instant of cosmic time. This property makes 
them suitable as model of our universe. The homogeneous and isotropic FRW cosmological models which 
are used to describe standard cosmological models, are particular case of Bianchi-V universes, according 
to whether the constant curvature of physical three-space, t = constant, is negative.\\

In this paper we have established the existence of string cosmological model with 
time varying DP in Bianchi-V space-time unlike the other authors. The organization of 
the paper is as follows: The model and field equations are presented in section 2. Section 3 deals 
with solution of field equations. The physical and geometrical properties of model is presented 
in section 4. Finally the conclusions of the paper are presented in section 5.   


\section{Bianchi V model and the field equations}
The line element for the spatially homogeneous and anisotropic Bianchi-V space-time is given by
\begin{equation}
 \label{eq1}
ds^{2}=-dt^{2}+A^{2}dx^{2}+e^{2\alpha x}\left(B^{2}dy^{2}+C^{2}dz^{2}\right)
\end{equation}
where $A(t)$, $B(t)$ and $C(t)$ are the scale factors in different spatial directions and 
$\alpha$ is a constant.\\

We define $a = (ABC)^{1/3}$ as the average scale factor of the space-time (\ref{eq1}) so that 
the average Hubble's parameter reads as
\begin{equation}
 \label{eq2}
H=\frac{\dot{a}}{a}
\end{equation}
where an over dot denotes derivative with respect to the cosmic time t.\\

The energy-momentum tensor $T^{i}_{j}$ for a cloud
of massive strings and perfect fluid distribution is taken as
\begin{equation}
\label{eq3} T^{i}_{j} = (\rho + p)v^{i}v_{j} + p g^{i}_{j} -\lambda
x^{i}x_{j},
\end{equation}
where $p$ is the isotropic pressure; $\rho$ is the proper energy density for a cloud strings with particles
attached to them; $\lambda$ is the string tension density; $v^{i}=(0,0,0,1)$ is the four-velocity of the
particles, and $x^{i}$ is a unit space-like vector representing the direction of string. The vectors $v^{i}$
and $x^{i}$ satisfy the conditions
\begin{equation}
\label{eq4} v_{i}v^{i}=-x_{i}x^{i}=-1,\;\; v^{i}x_{i}=0.
\end{equation}

Choosing $x^{i}$ parallel to $\partial/\partial x$, we have
\begin{equation}
\label{eq5} x^{i} = (A^{-1},0,0,0).
\end{equation}
If the particle density of the configuration is denoted by
$\rho_{p}$, then
\begin{equation}
\label{eq6} \rho = \rho_{p}+\lambda.
\end{equation}
The Einstein's field equations (in gravitational units $c = 1$, $8\pi G = 1$) read as
\begin{equation}
\label{eq7}
R_{j}^i - \frac{1}{2}g_{j}^{i}R  = -T_{j}^i
\end{equation}
The Einstein's field equations (\ref{eq7}) for the line-element (\ref{eq1}) 
lead to the following system of equations 
\begin{equation}
 \label{eq8}
\frac{\ddot{B}}{B}+\frac{\ddot{C}}{C}+\frac{\dot{B}\dot{C}}{BC}-\frac{\alpha^{2}}{A^{2}} = -p + \lambda
\end{equation}
\begin{equation}
 \label{eq9}
\frac{\ddot{A}}{A}+\frac{\ddot{C}}{C}+\frac{\dot{A}\dot{C}}{AC}-\frac{\alpha^{2}}{A^{2}} = -p
\end{equation}
\begin{equation}
 \label{eq10}
\frac{\ddot{A}}{A}+\frac{\ddot{B}}{B}+\frac{\dot{A}\dot{B}}{AB}-\frac{\alpha^{2}}{A^{2}} = -p
\end{equation}
\begin{equation}
 \label{eq11}
\frac{\dot{A}\dot{B}}{AB}+\frac{\dot{A}\dot{C}}{AC}+\frac{\dot{B}\dot{C}}{BC}-\frac{3\alpha^{2}}{A^{2}} = \rho
\end{equation}
\begin{equation}
 \label{eq12}
\frac{2\dot{A}}{A}-\frac{\dot{B}}{B}-\frac{\dot{C}}{C} = 0
\end{equation}
The energy conservation equation $T^{i}_{j\;;j} = 0$ leads to the following expression
\begin{equation}
 \label{eq13}
\dot{\rho}+(\rho +p)\left(\frac{\dot{A}}{A}+\frac{\dot{B}}{B}+\frac{\dot{C}}{C}\right)-\lambda\frac{\dot{A}}{A}=0
\end{equation}
which is a consequence of the field equations.\\
\section{Solution of field equations}
Integrating equation (\ref{eq12}) and absorbing the constant of integration in $B$ or $C$, without 
loss of generality, we obtain
\begin{equation}
 \label{eq14}
A^{2} = BC
\end{equation}
Subtracting equation (\ref{eq9}) from (\ref{eq10}) and taking second integral, we 
get the following relation
\begin{equation}
 \label{eq15}
\frac{B}{C}=d_{1}exp\left(x_{1}\int{\frac{dt}{ABC}}\right)
\end{equation}
where $d_{1}$ and $x_{1}$ are the constant of integrations.\\
Equations (\ref{eq8})$-$(\ref{eq12}) are five independent equations in six unknown $A$, $B$, $C$, $p$, $\rho$ and 
$\lambda$. For the complete determination of the system, we need one extra condition.\\
Following Pradhan and Amirhashchi (2011),we assume that
\begin{equation}
 \label{eq16}
a=\left(t^{n}e^{t}\right)^{\frac{1}{m}}
\end{equation}
where $m$ and $n$ are positive constants. It is important to note here that 
the ansatz for scale factor generalized the one proposed by 
Pradhan and Amirhashchi (2011) because for $m = 2$, we obtain the same 
expression for $a$ as proposed by Pradhan and Amirhashchi (2011) to 
describe the dark energy model in Bianchi V space-time. However in this paper, we considered string 
fluid as source of matter.\\
Now the spatial volume $V$ of the model read as
\begin{equation}
 \label{eq17}
V=a^{3}=\left(t^{n}e^{t}\right)^{\frac{3}{m}}
\end{equation}
Equations (\ref{eq14}), (\ref{eq16}) and (\ref{eq17}) lead to
\begin{equation}
 \label{eq18}
A(t)=\left(t^{n}e^{t}\right)^{\frac{1}{m}}
\end{equation}
Inserting (\ref{eq18}) into (\ref{eq14}) and (\ref{eq15}), we get
\begin{equation}
 \label{eq19}
B=\left(t^{n}e^{t}\right)^{\frac{1}{m}}\sqrt{d_{1}}exp\left(\frac{x_{1}}{2}\int{\frac{dt}
{\left(t^{n}e^{t}\right)^{\frac{3}{m}}}}\right)
\end{equation}
\begin{equation}
 \label{eq20}
C=\frac{\left(t^{n}e^{t}\right)^{\frac{1}{m}}}{\sqrt{d_{1}}}exp\left(-\frac{x_{1}}{2}\int{\frac{dt}
{\left(t^{n}e^{t}\right)^{\frac{3}{m}}}}\right)
\end{equation}
\section{Some physical and geometrical properties}
The isotropic pressure $(p)$, proper energy density $(\rho)$, string tension density $(\lambda)$ and 
particle density $(\rho_{p})$ are given by
\begin{equation}
 \label{eq21}
p=\alpha^{2}(t^{n}e^{t})^{-\frac{2}{m}}+\frac{x_{1}}{m}\left(\frac{n}{t}+1\right)(t^{n}e^{t})^{-\frac{3}{m}}-
\left[\frac{1}{m}\left(\frac{n}{t}+1\right)+\frac{x_{1}}{2}(t^{n}e^{t})^{-\frac{3}{m}}\right]^{2}
\end{equation}
\begin{equation}
\label{eq22}
\rho=\frac{3}{m^{2}}\left(\frac{n}{t}+1\right)^{2}-\frac{x_{1}^{2}}{4}(t^{n}e^{t})^{-\frac{6}{m}}-
3\alpha^{2}(t^{n}e^{t})^{-\frac{2}{m}}
\end{equation}
\begin{equation}
 \label{eq23}
\lambda=\frac{x_{1}}{m}\left(\frac{n}{t}+1\right)(t^{n}e^{t})^{-\frac{3}{m}}+\frac{1}{m^{2}}\left(\frac{n}{t}+1\right)^{2}
+\left[\frac{1}{m}\left(\frac{n}{t}+1\right)+\frac{x_{1}}{2}(t^{n}e^{t})^{-\frac{3}{m}}\right]^{2}-\frac{2n}{mt^{2}}-
\frac{x_{1}^{2}}{4}(t^{n}e^{t})^{-\frac{6}{m}}
\end{equation}
\begin{equation}
 \label{eq24}
\rho_{p}=\frac{2}{m^{2}}\left(\frac{n}{t}+1\right)^{2}+\frac{2n}{mt^{2}}-\frac{x_{1}}{m}\left(\frac{n}{t}+1\right)
(t^{n}e^{t})^{-\frac{3}{m}}-
\left[\frac{1}{m}\left(\frac{n}{t}+1\right)+\frac{x_{1}}{2}(t^{n}e^{t})^{-\frac{3}{m}}\right]^{2}
-3\alpha^{2}(t^{n}e^{t})^{-\frac{2}{m}}
\end{equation}
The average Hubble's parameter $(H)$, expansion scalar $(\theta)$, anisotropy parameter $(A_{m})$ and 
shear scalar $(\sigma)$ of the model are given by
\begin{equation}
 \label{eq25}
H=\frac{1}{m}\left(\frac{n}{t}+1\right)
\end{equation}
\begin{equation}
 \label{eq26}
\theta=3H=\frac{3}{m}\left(\frac{n}{t}+1\right)
\end{equation}
\begin{equation}
 \label{eq27}
A_{m}=\frac{1}{9H^{2}}\left[\left(\frac{\dot{A}}{A}-\frac{\dot{B}}{B}\right)^{2}+
\left(\frac{\dot{B}}{B}-\frac{\dot{C}}{C}\right)^{2}+\left(\frac{\dot{C}}{C}-\frac{\dot{A}}{A}\right)^{2}\right]=
\frac{m^{2}x_{1}^{2}}{6\left(\frac{n}{t}+1\right)}(t^{n}e^{t})^{-\frac{6}{m}}
\end{equation}
\begin{equation}
 \label{eq28}
\sigma^{2}=\frac{3}{2}A_{m}H^{2}=\frac{x_{1}^{2}}{4(t^{n}e^{t})^{\frac{6}{m}}}
\end{equation}
The value of DP $(q)$ is found to be
\begin{equation}
 \label{eq29}
q=-1+\frac{mn}{(n+t)^{2}}
\end{equation}
\begin{figure}
\begin{center}
\includegraphics[width=4.0in]{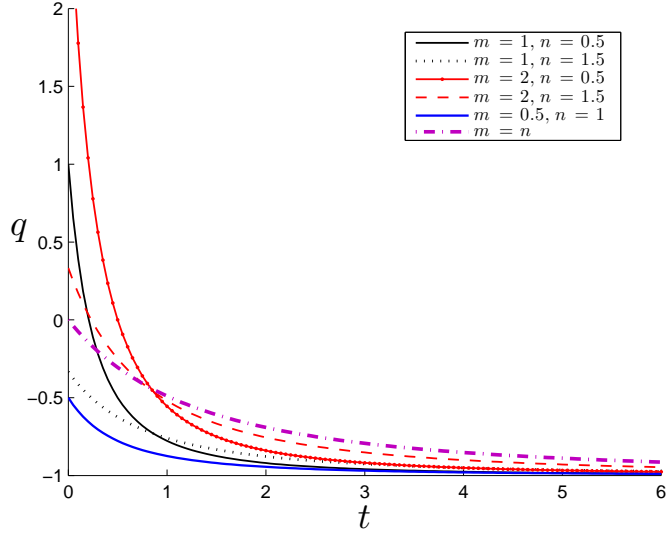} 
\caption{The plot of DP $(q)$ vs. time (t).}
\label{fg:anil37F1.eps}
\end{center}
\end{figure}
From Fig. 1, we see that the dynamics of DP $(q)$ depends on two free parameters $m$ and
$n$. For $m = n$ (violet dash-dotted line), $m = 1$ and $n = 1.5$ (black dotted line) and 
$m = 0.5,\; n = 1$ (blue solid line), the universe is accelerating from its birth whereas for 
$m = 1,\; n = 0.5$ (solid black line), $m = 2,\; n = 0.5$ (solid red line with dot marker) 
and $m = 2,\; n = 1.5$ (dashed red line), it goes to transition from early decelerating 
phase to current accelerating phase. Since, we are looking for a model, describing the universe from early 
decelerating phase to current accelerating phase hence we choose $m = 1$ and 
$n = 0.5$ in the remaining discussions of the model as it is the most appropriate 
choice in our case.\\

From equation (\ref{eq29}), it is clear that the DP $(q)$ is time dependent. 
Also, the transition redshift from deceleration expansion to accelerated expansion 
is about $0.5$. Now for a universe which was decelerating in past and accelerating at 
present time, DP must show signature flipping (Amendola 2003; Riess et al. 2001).\\
 
\begin{figure}
\begin{center}
\includegraphics[width=4.0in]{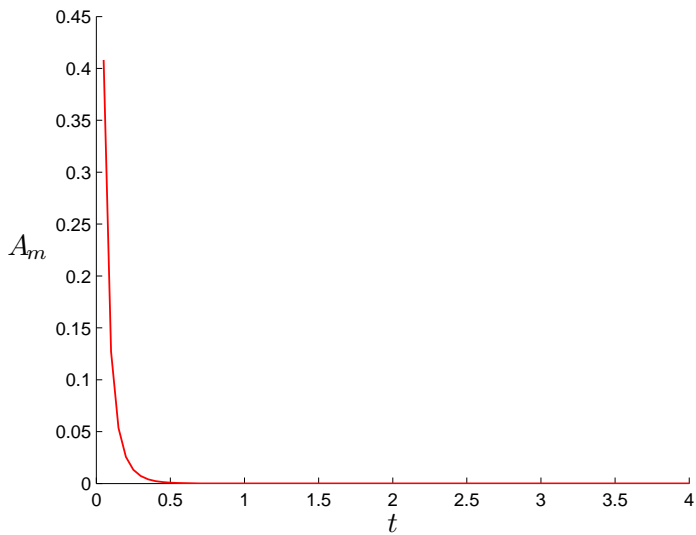} 
\caption{The plot of anisotropic parameter $A_{m}$ vs. time (t).}
\label{fg:anil37F2.eps}
\end{center}
\end{figure}

\begin{figure}
\begin{center}
\includegraphics[width=4.0in]{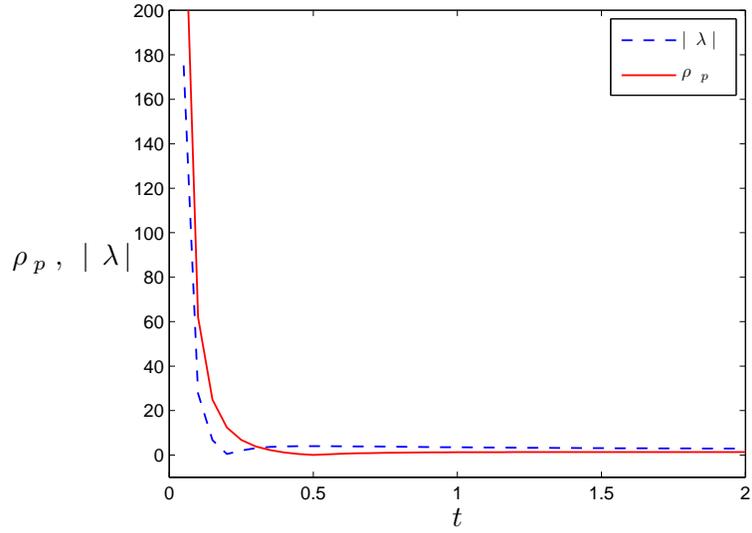} 
\caption{The particle density $(\rho_{p})$ and string tension density $(\lambda)$ vs. time (t).}
\label{fg:anil37F3.eps}
\end{center}
\end{figure}

It is observed that at $t = 0$, the spatial volume vanishes 
and other parameters $H$, $\theta$ and $\sigma$ diverge. Hence the model 
starts with big bang singularity at $t = 0$ and this singularity is point type because 
all directional scale factors vanish at initial moment.
Figure 2 depicts the variation of anisotropic parameter $(A_{m})$ 
versus cosmic time. It is shown that $A_{m}$ decreases with time and tends to zero for 
sufficiently large times. Thus the anisotropic behaviour of universe dies out on later times 
and the observed isotropy of universe can be achieved in derived model at present epoch.\\

In this model the universe starts with finite values of proper energy density $(\rho)$, 
particle energy density $(\rho_{p})$ and string tension density $(\lambda)$ but with evolution of 
universe, $\rho$ and $\lambda$ becomes negligible for sufficiently long time (i. e. at present epoch). 
The large value of $\rho$ and $\arrowvert \lambda \arrowvert$ 
in the beginning suggest that string dominates the early universe but on later times, 
$\rho_{p}$ and $\arrowvert \lambda \arrowvert$ becomes negligible. Thus the strings 
disappear from the universe for larger times and hence it is not observable today. This 
behaviour of $\rho_{p}$ and $\arrowvert \lambda \arrowvert$ is shown in Figure 3. Since there is 
no direct evidence of strings in present day universe, we are in general, interested in constructing 
model of a universe that evolves purely from the era dominated by either geometric strings or massive 
strings and end up in a particle dominated era with or without remnants of strings. Therefore, the above 
model describes the evolution of the universe consistent with the present day observations.

\begin{figure}
\begin{center}
\includegraphics[width=6.0in]{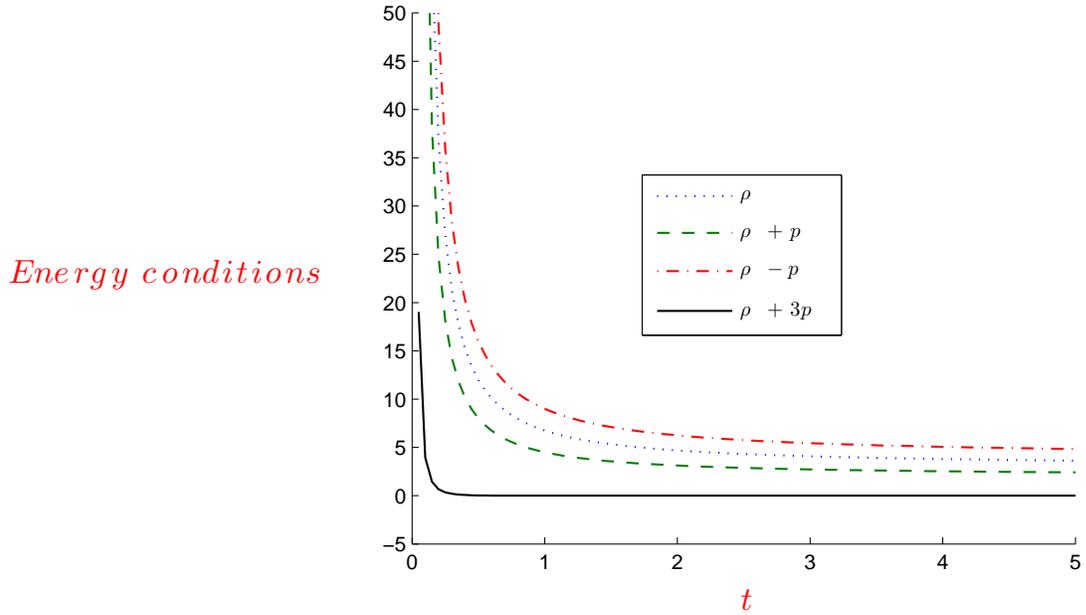} 
\caption{The energy conditions vs. time (t).}
\label{fg:anil37F4.eps}
\end{center}
\end{figure}
The left hand side of energy conditions have graphed in Figure 4. From Figure 4, we observe that\\
(i)\; $\rho > 0$\;\;\\
(ii)\; $\rho + p > 0$\;\;\\
(iii)\; $\rho - p > 0$ \\

Therefore weak energy condition (WEC) as well as dominant energy condition (DEC) are 
satisfied in the derived model. It is also observed that $\rho + 3p > 0$ at initial moment 
and on later times $\rho + 3p \leq 0$ which in turn imply that the strong energy condition (SEC) 
violates in derived model on later times. The violation of SEC gives a reverse 
gravitational effect. Due to this effect, the universe gets jerk and the transition from 
the earlier deceleration phase to the recent acceleration phase take place (Caldwell et al. 2006). 
Thus the model presented in this paper 
is turning out as a suitable model for describing the late time acceleration of the universe.\\

\section{Concluding Remarks}
In this paper, a spatially homogeneous and anisotropic Bianchi-V string model has been investigated for 
which the string fluid are rotation free but they do have expansion and shear. We observe that 
$V \rightarrow \infty$ and $\rho \rightarrow 0$ as $t\rightarrow \infty$ \;i. e. spatial volume $(V)$ increases with 
time and proper energy density $(\rho)$ decreases with time as expected. 
The main features of the derived model are as follows:

\begin{itemize}
 
\item
The model is based on exact solution of Einstein's field equations for 
the anisotropic Bianchi-V space-time filled with string fluid as source of matter.

\item
The dynamics of DP yields two different phases of the universe. Initially DP is evolving with 
positive sign that yields the decelerating phase of the universe whereas in later times, it is evolving 
with negative sign which describes the present phase of acceleration of the universe. Thus the derived 
model has transition of the universe from early decelerating phase to the current accelerating phase.

\item 
In the present model, the WEC and DEC are satisfied which in turn imply that the derived 
models are physically realistic while the violation of SEC is in agreement with current 
astrophysical observations.

\item
The string tension density $(\lambda)$ start off with extremely large values and 
tends to zero for sufficiently large times in derived model. Thus the strings dominates the early universe 
and eventually disappear from the universe at later times (i. e. present 
epoch).     
 
\end{itemize}

Finally, the solution presented here can be one of the potential candidates to describe the 
observed universe. Therefore physically viable Bianchi-V string cosmological model with 
singular origin has been obtained.\\

\subsection*{Acknowledgements}
Author would like to thanks the anonymous referee for his valuable suggestions to improve 
the quality of this manuscript.

\end{document}